\documentclass[aps,prl,twocolumn,superscriptaddress,floatfix,showpacs]{revtex4}

\usepackage{graphics}
\usepackage{epsfig}
\usepackage{amsmath}
\usepackage{color}
\usepackage{verbatim}

\newcommand{\apsi}
{\affiliation{Paul Scherrer Institute, CH-5232 Villigen PSI, Switzerland}}
\newcommand{\aucb}
{\affiliation{Department of Physics, University of California, Berkeley, and LBNL, Berkeley, CA 94720, USA}}
\newcommand{\apnpi}
{\affiliation{Petersburg Nuclear Physics Institute, Gatchina 188350, Russia}}
\newcommand{\auiuc}
{\affiliation{Department of Physics, University of Illinois at Urbana-Champaign, Urbana, IL 61801, USA}}
\newcommand{\auw}
{\affiliation{Department of Physics, University of Washington, Seattle, WA 98195, USA}}
\newcommand{\aucl}
{\affiliation{Institute of Nuclear Physics, Universit{\'e} Catholique de Louvain, B-1348, Louvain-la-Neuve, Belgium}}
\newcommand{\auk}{\affiliation{Department of Physics and Astronomy, University of Kentucky, Lexington, KY 40506, USA}}
\newcommand{\aub}{\affiliation{Department of Physics, Boston University, Boston, MA 02215, USA}}
\newcommand{\aregis}
{\affiliation{Department of Physics and Computational Science, Regis University, Denver, CO 80221, USA}}

\newcommand{\is}{s$^{-1}\;$}
\newcommand{\isn}{s$^{-1}$}
\newcommand{\mathis}{\textrm{s}^{-1}}

\newcommand{\LS}{$\Lambda_{S}$}
\newcommand{\gp}{$g^{}_{P}$}

\newcommand{\gpm}{g^{}_{P}}

\newcommand{\LambdaMuplus}{455170.05}
\newcommand{\LambdaMuplusErr}{0.46}

\newcommand{\LambdaSEight}{713.5} 
\newcommand{\LambdaSStatEight}{12.5}
\newcommand{\LambdaSSystEight}{8.6}
 
\newcommand{\LambdaSTen}{717.3}
\newcommand{\LambdaSStatTen}{7.73}
\newcommand{\LambdaSSystTen}{5.55}

\newcommand{\LambdaSEleven}{713.1}
\newcommand{\LambdaSStatEleven}{8.33}
\newcommand{\LambdaSSystEleven}{4.34}

\newcommand{\LambdaSThErrRad}{3.0}
\newcommand{\LambdaS}{714.9}
\newcommand{\LambdaSStat}{5.4}
\newcommand{\LambdaSSyst}{5.1}

\newcommand{\gPth}{8.26}
\newcommand{\gPtherr}{0.23}

\begin{document}

\title{Measurement of Muon Capture on the Proton to 1\% Precision and\\ 
Determination of the Pseudoscalar Coupling $g^{}_P$}
\author{V.A.~Andreev}
\apnpi
\author{T.I.~Banks}
\aucb
\author{R.M.~Carey}
\aub
\author{T.A.~Case}
\aucb
\author{S.M.~Clayton}
\auiuc
\author{K.M.~Crowe\footnote{Deceased}}
\aucb
\author{J.~Deutsch\footnotemark[\value{footnote}]}
\aucl
\author{J.~Egger}
\apsi
\author{S.J.~Freedman\footnotemark[\value{footnote}]}
\aucb
\author{V.A.~Ganzha}
\apnpi
\author{T.~Gorringe}
\auk
\author{F.E.~Gray}
\aregis
\aucb
\author{D.W.~Hertzog}
\auiuc \auw
\author{M.~Hildebrandt}
\apsi
\author{P.~Kammel}
\auiuc \auw
\author{B.~Kiburg}
\auiuc \auw
\author{S.~Knaack}
\auiuc
\author{P.A.~Kravtsov}
\apnpi
\author{A.G.~Krivshich}
\apnpi
\author{B.~Lauss}
\apsi
\author{K.R.~Lynch}
\aub
\author{E.M.~Maev}
\apnpi
\author{O.E.~Maev}
\apnpi
\author{F.~Mulhauser}
\auiuc
\apsi
\author{C.~Petitjean}
\apsi
\author{G.E.~Petrov}
\apnpi
\author{R.~Prieels}
\aucl
\author{G.N.~Schapkin}
\apnpi
\author{G.G.~Semenchuk}
\apnpi
\author{M.A.~Soroka}
\apnpi
\author{V.~Tishchenko}
\auk
\author{A.A.~Vasilyev}
\apnpi
\author{A.A.~Vorobyov}
\apnpi
\author{M.E.~Vznuzdaev} 
\apnpi
\author{P.~Winter}
\auiuc \auw
\collaboration{MuCap Collaboration}
\date{\today}

\begin{abstract}
The MuCap experiment at the Paul Scherrer Institute has measured the rate \LS\ of muon capture from the singlet state of the muonic 
hydrogen atom to a precision of 1\,\%. A muon beam was stopped in a time projection chamber filled with 10-bar, ultrapure hydrogen gas. 
Cylindrical wire chambers and a segmented scintillator barrel detected electrons from muon decay. 
 \LS\ is determined from the difference between the $\mu^-$ disappearance rate in hydrogen and the free muon decay rate.
The result is based on the analysis of $1.2\times10^{10}$ $\mu^-$ decays, from which we extract the capture rate 
$\Lambda_{S}^{} = (\LambdaS \pm \LambdaSStat_{\textrm{stat}} \pm
\LambdaSSyst_{\textrm{syst}})$ \is and derive the proton's pseudoscalar coupling $g_{P}^{}(q_0^2 = -0.88\,m_\mu^2) = 8.06 \pm 0.55$. 
\end{abstract}
\maketitle

We report a measurement of the rate \LS\ of ordinary muon capture (OMC),
\begin{equation}
\mu^{-} + p \rightarrow n + \nu_{\mu} \label{equation:omc},
\end{equation}
from the singlet state of the muonic hydrogen atom. 
The analysis uses the complete data set of the MuCap experiment, with significantly smaller systematic and statistical uncertainties compared to our earlier publication~\cite{Andreev:2007}.

For the low momentum transfer $q_0^2 = -0.88m_\mu^2$ in process
\eqref{equation:omc}, the standard model electroweak interaction
reduces to an effective Fermi interaction between the leptonic and
hadronic weak currents. While the leptonic current has a simple
$\gamma_\mu (1-\gamma_5)$  structure, the hadronic current between
nucleon states is modified by QCD, as expressed in a model-independent 
way by the introduction of form factors. Since second-class currents are suppressed, muon
capture on the proton involves $g^{}_V(q_0^2)$ and $g^{}_M(q_0^2)$,
the vector and magnetic form factors in the vector current, as well as
$g^{}_A(q_0^2)$ and $g^{}_P(q_0^2)$,  the  axial and pseudoscalar form
factors in the axial 
current~\cite{Bernard:2001rs, Gorringe:2002xx,  Kammel:2010}. 
The first three are well known and contribute only around 0.4\% uncertainty 
to the determination of
\LS~\cite{Beringer:1900zz}. Our measurement of \LS\ determines
$\gpm \equiv \gpm(q_0^2)$, the least well known of these form factors. 

The pseudoscalar term in the axial nucleon current has played a significant 
role in the understanding of weak and strong interactions. Initial estimates 
were based on the concept of a partially conserved axial current, followed 
by the recognition of its deeper significance as a consequence of chiral 
symmetry and its spontaneous and explicit 
breaking~\cite{Nambu:1960xd}. 
These ideas were foundations for explaining the generation of hadronic 
masses and the development of chiral perturbation theory (ChPT), the 
effective field theory of low-energy QCD.
 Based on well-known low-energy constants,
\begin{equation}
g_{P}^{\textrm{theory}}=\gPth \pm \gPtherr
 \label{equation:gp_theory}
\end{equation}
was derived within ChPT~\cite{Bernard:2001rs,Bernard:1994wn}, with
good convergence to two-loop order~\cite{Kaiser:2003dr}. Though lattice QCD has advanced to unquenched calculations of $g^{}_A$ and $g^{}_P$~\cite{Yamazaki:2009, Alexandrou:2011}, the precision of the ChPT prediction in Eqn.~\ref{equation:gp_theory} remains unmatched and stands to be tested experimentally.

Muon capture on hydrogen is the most direct means to 
determine \gp. Such experiments are complicated by the fact that 
negative muons stopped in hydrogen form not only $\mu p$ atoms, 
but subsequently $pp\mu$ molecules where the capture rate differs 
significantly.
Prior to MuCap, the most precise capture rate was measured in liquid hydrogen~(LH$_2$)~\cite{Bardin:1980mi}, where $pp\mu$ forms rapidly. The value of \gp\ 
extracted under these conditions depends critically on the poorly known 
ortho-to-para $pp\mu$ transition rate, 
$\lambda_{op}$~\cite{Bardin:1981cq,Clark:2005as}. The related, rare 
radiative muon capture (RMC) process $\mu^{-} p \rightarrow n  \nu \gamma$ 
is less sensitive to  $\lambda_{op}$, but the first experimental
result on \gp~\cite{Wright:1998gi} disagreed with theory. For a discussion 
of this puzzling situation and muonic processes in hydrogen, see~\cite{Bernard:2001rs, Gorringe:2002xx,  Kammel:2010}.

\begin{figure}
\begin{center}
\includegraphics[width=0.49\textwidth]{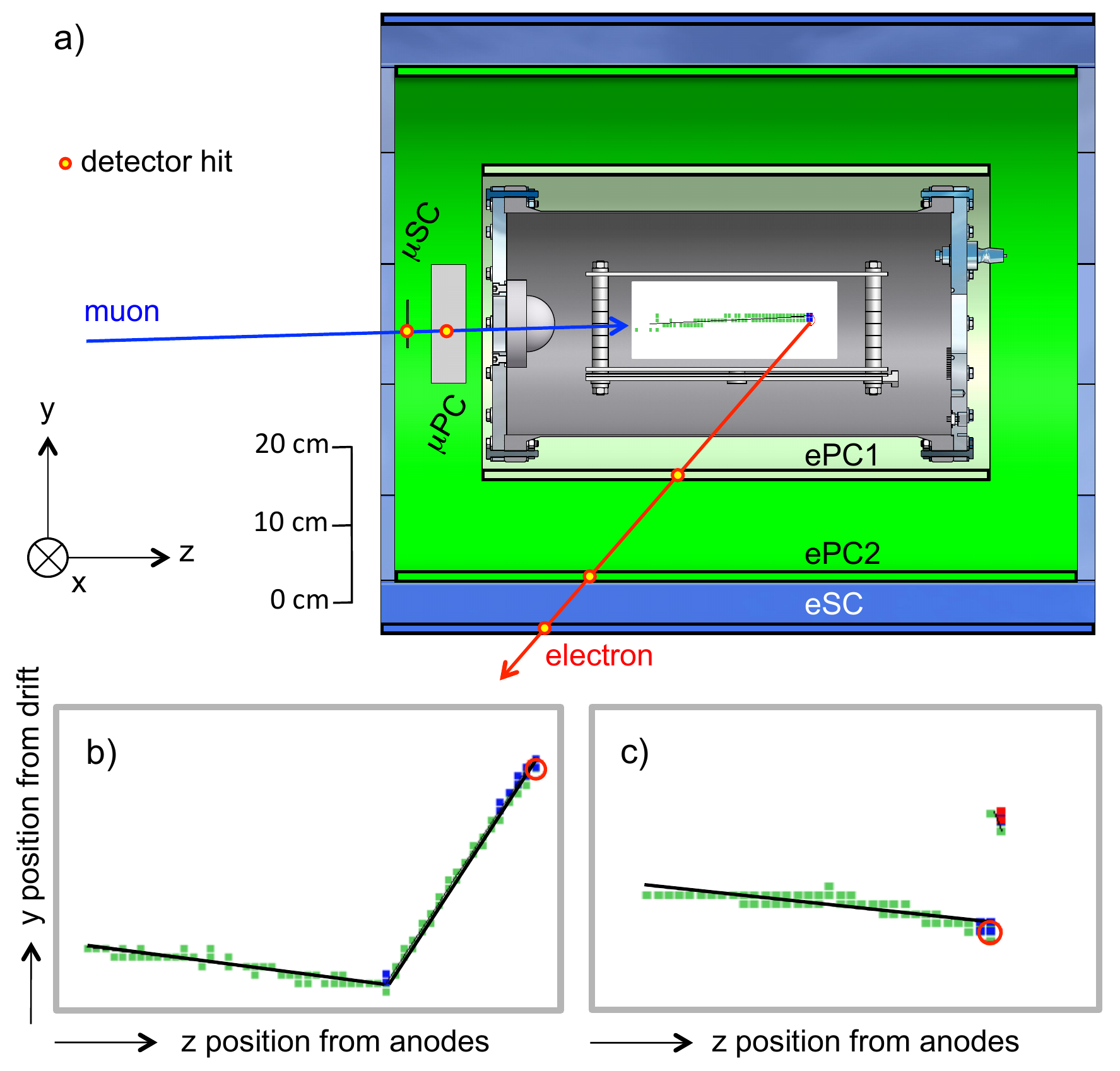}
\end{center}
\vspace{-5mm}
  \caption{(color online)  Upper panel: (a) Cross-sectional view of the MuCap detector showing a typical muon stop and decay electron. Lower panels: Zoomed-in special event topologies. (b) Rare large-angle $\mu$-$p$ scatter event with $\sim 10^{-2}/\mu$ probability. c) Very rare delayed capture on impurity in $\mu O \rightarrow N^* \nu $ with $\sim 3.4\times10^{-6}/\mu$ probability. The interpretation of the event displays is described in the text.}
  \label{evd.fig}
\vspace{-5mm}
\end{figure}
 The MuCap experiment was designed to significantly reduce the density-dependent formation of $pp\mu$ molecules by employing a gas density of  $\phi \approx 0.01$ (relative to LH$_2$). In these conditions about 97\% of the muon captures occur in the $\mu p$ singlet state. The experimental concept is sketched in Figure \ref{evd.fig}(a). A 34~MeV/$c$ 
muon beam was stopped in a time projection chamber (TPC) filled with 10-bar, 
ultrapure hydrogen gas \cite{Egger:2011zz}. The TPC was used to discriminate 
between muons that stop in the gas and those that reach wall materials, where 
capture proceeds much faster than in hydrogen. Arriving muons were detected 
by an entrance scintillator ($\mu$SC) and a proportional chamber ($\mu$PC), 
and tracked in the TPC. Outgoing decay electrons were detected by concentric 
multiwire proportional chambers (ePC1 and ePC2) and a segmented scintillator 
barrel (eSC). The decay times were histogrammed and fit to an
exponential. The difference between the observed disappearance rate 
$ \lambda_{\mu^{-}}$ and the free muon decay rate 
$\lambda_{\mu^{+}}$ \cite{Webber:2011} is attributed to muon capture, 
$\Lambda_{S} \approx \lambda_{\mu^{-}} - \lambda_{\mu^{+}}$.  

The new data reported here were collected during the 2006 (R06)
and 2007 (R07) running periods in the $\pi$E3 muon channel at the Paul 
Scherrer Institute. 
Properties of the new data sets are compared to our published result 
(R04)~\cite{Andreev:2007} in Table~\ref{tbl:RunTable}. Besides the $1.2\times10^{10}$ muon-electron pairs from $\mu^-$ stops in hydrogen, additional systematic data  included $0.6 \times 10^{10}$ $\mu^+$ decays and $\mu^-$ data collected when the target gas was doped with elemental impurities (nitrogen, water and argon).

\begin{table}[ht!]
  \centering
  \caption{Main features of MuCap production runs. Statistics of fully reconstructed $\mu$-$e$ pairs, deuterium concentration $c^{}_D$, water concentration $c^{}_{H_2O}$ determined by humidity sensor (not present in R04), and observed impurity capture yield per muon, $Y_Z$.\label{tbl:RunTable}}
  \begin{tabular}{lccc}
    \hline
    Quantity       &  R04  &  R06   &  R07 \\
    \hline
    Statistics &  $1.6 \times 10^9$  & $5.5\times 10^9$  & $5.0 \times 10^9$\\
    $c^{}_{D}$ (ppb) & 1440   & $<$60  & $<$10 \\
    $c^{}_{H_2O}$ (ppb) &  $\cdots$   & 18.  & 8.7 \\
    $Y^{}_{Z}$ (ppm)   & 12 & 6.3 & 3.4 \\
    \hline
 \end{tabular}
 \end{table}

Although the experimental methodology closely followed that of our first result, several hardware upgrades implemented between R04 and R06 led to significantly enhanced performance. In R06 the TPC was operated with about 2.5 times higher gas gain than in R07. As this affects critical chamber parameters, the comparison of the two runs provides an invaluable consistency check.

Events with multiple muons in the TPC (pileup) need to be rejected as they distort the extracted disappearance rate. The maximum rate of the dc muon beam employed in R04 was throttled to minimize pileup. In R06 and R07, the loss of events to pileup
was largely eliminated by the introduction of a 25-kV, fast-switching 
electrostatic kicker \cite{Barnes20041932}.
The detection of a muon traversing the $\mu$SC triggered the kicker,
which deflected the beam for a period of 25.6 $\mu$s. The beam extinction factor
was around 100 and the rate of pileup-free data was three times larger 
in R06/7 than in R04.

Another essential change was the greatly improved isotopic and chemical 
purity of the TPC gas.  When deuterium is present (concentration~$c^{}_D$), 
muons can form $\mu d$ atoms, which, due to a Ramsauer-Townsend minimum in the scattering cross section~\cite{Kammel:2010}, can diffuse out of the fiducial volume and distort the disappearance rate. To separate 
the hydrogen into its isotopic components, a new cryogenic distillation column 
was installed. Periodic gas samples were analyzed
externally using atomic mass spectrometry~\cite{2008NIMPB.266.1820D}. 
For the limits on $c^{}_D$ listed in Table~\ref{tbl:RunTable}, 
transfer to $\mu d$ leads to distortions of less than 0.74~\is and 0.12~\is for the
R06 and R07 run periods, respectively. A higher sensitivity of the atomic mass spectrometer was responsible for the improved limit in R07.

In the presence of $Z>1$ impurities, 
muons preferentially transfer from $\mu p$ to $\mu Z$ atoms, distorting 
the disappearance rate. Extended baking of the TPC and increased flux and 
filtering in the continuous gas purification system~\cite{Ganzha:2007uk} 
led to a fourfold reduced impurity level compared to R04. Moreover, the 
installation of a humidity sensor before R06 allowed monitoring of the dominant chemical impurity. 
	
Our novel hydrogen TPC was key to the experiment. In the sensitive volume of 
$x \times y \times z = 15\times12\times28$~cm$^{3}$ with an applied field of 2 kV/cm, ionization electrons from stopping muons drifted down towards the readout plane with a velocity $v_y= -5.5\textrm{ mm/}\mu$s. They were 
amplified in a multiwire proportional chamber region with 72 anodes 
perpendicular and 36 strip wires parallel to the beam. 
Anode and cathode signals were discriminated 
with three energy thresholds and read out by TDCs in 200\,ns time 
intervals. The event display in the center of Figure \ref{evd.fig}(a) 
shows the $y$-$z$ projection of a typical muon stop in 
the TPC. The threshold $E_1\approx15$~keV is indicated by green pixels.  
Blue pixels denote the threshold $E_2\approx55$~keV, which was set just 
below the muon's Bragg peak. The threshold $E_3\approx315$~keV [red 
pixels in Figure \ref{evd.fig}(c)] was set to record nuclear recoils 
from muon captures on impurities. In addition to the primary TDC-based readout of the TPC, new 12-bit, 25\,MHz flash analog-to-digital 
converters recorded selectively triggered events.

About 30\,TB of raw data were processed at the NCSA supercomputing facility 
in a multistage procedure. Muon stop candidates were constructed from the 
$\mu$SC time and TPC track information. Muon pileup events, flagged by the 
entrance counters, were rejected. Because the combined inefficiency of the 
entrance counters was less than $10^{-4}$, the residual pileup 
distorted the observed muon 
disappearance rate $\lambda_{\mu^{-}}$ by less than 0.5\,\isn. Contiguous 
pixel regions in the TPC were then fit to a straight line, as indicated 
by the black line 
along the muon trajectory in Fig. \ref{evd.fig}(a). In the case of large 
angle scattering [Fig.~\ref{evd.fig}(b)] a two-line fit was applied. The 
muon stop location (red circle) was identified as the most downstream 
$E_2$ pixel. The muon track requirements were optimized 
so as to minimize possible distortions to $\lambda_{\mu^{-}}$ 
while suppressing events where the muon could have left the hydrogen 
gas. Muons that stopped within a fiducial volume 
$\Delta x \times \Delta y \times \Delta z = 10.4\times8.0\times20.4\,\textrm{cm}^{3}$ were accepted. The minimum track length was 3.2~cm, and the maximum fit $\chi^2$ was 2.
$\lambda_{\mu^{-}}$ was stable against variation of the track length 
and $\chi^2$ cuts. However, variations in the fiducial volume boundaries 
produced statistically disallowed deviations, for which a systematic 
uncertainty of $3.0$\,\isn ~was assigned.

Electron tracks were constructed from coincidences between an eSC segment 
(comprising four photomultiplier tubes) and hits in the two ePCs (each requiring an anode and at least one cathode plane). 
In the R06 and R07 run periods, the time and gain stabilities of the eSC  
were verified by recording their signals in 8-bit, 450\,MHz
waveform digitizers. While the TPC gain was insufficient to produce  
electron tracks with contiguous pixels, 
a virtual track in the TPC was reconstructed from hits in the
eSC and ePCs, as indicated by the red line in Figure~\ref{evd.fig}(a). A cut of $b \leq120$~mm was
placed on the impact parameter $b$ between the muon stop and electron vector. 
This loose cut significantly reduces backgrounds (c.f. Fig.~3 in~\cite{Andreev:2007}) while minimizing distortions of $\lambda_{\mu^{-}}$ introduced by
a time-dependent acceptance due to $\mu p$ diffusion. Although $\mu p$ atoms
diffuse only at the mm scale, changes in $\lambda_{\mu^{-}}$ 
vs. $b$ were observed. This $\lambda_{\mu^{-}}(b)$ dependence was used to fixed the single parameter of a $\mu p$ diffusion model in good agreement with theory~\cite{adamczak:042718}. For the applied cut, the model 
was used to determine small corrections $(-3.1\pm0.1)$\,\is  
and $(-3.0\pm0.1)$\,\is for R06 and R07, respectively. 
To check that $\lambda_{\mu^{-}}$  was insensitive to the electron track 
definition, we also constructed 
coincidences requiring different combinations of anode and cathode planes 
within the ePCs. 
This revealed slightly nonstatistical variations in  $\lambda_{\mu^{-}}$, which were fully covered by a 1.8\,\is systematic uncertainty.

In extreme cases of example Fig.~\ref{evd.fig}(b), muons scatter through large angles, leave the TPC volume, and 
stop on surrounding materials. Because of the lower TPC gain during R07, there were often gaps in the
tracks of scattered muons, making it difficult to reliably identify these events.
Moreover, the recoil proton could deposit enough energy in the TPC to 
trigger the $E_2$ threshold, mimicking an acceptable muon stop. 
However, these events were unlikely to deposit enough energy at the 
scattering vertex to exceed the $E_2$-threshold on neighboring anodes. 
In the analysis of the R06 and R07 data sets, we 
required at least two 
consecutive $E_2$ anodes at the end of the muon track. 

This cut introduced a subtle systematic effect. Electrons that 
traversed the muon's drifting ionization charge occasionally deposited 
enough additional energy to elevate a muon's $E_1$ signal above the $E_2$ 
threshold.  In rare instances a muon stop 
with a single $E_2$ anode was promoted to a stop with two neighboring $E_2$ 
anodes. Such events would pass the $\mu$-$p$ scatter cut described 
above with a decay-time-dependent acceptance and therefore distort the extracted disappearance rate. Because positive muons are sensitive to the charge interference 
effect but do not capture on nuclei, we were able to measure the induced distortion~\cite{kiburg}. 
The method was supplemented by neutron data collected in 8 large liquid
scintillator detectors: muons scattered into $Z>1$ materials were found to yield
copious neutrons from nuclear capture. The resulting corrections were $(-12.4 \pm 3.22)\,$\is and $(-7.2 \pm 1.25)\,$\is for R06 and R07, respectively. The correction was sensitive to the $E_2$ threshold, which we decreased in R07, suppressing the interference effect.

As illustrated in Fig. \ref{evd.fig}(c),  nuclear capture on impurities was 
identified by the presence of an $E_3$ threshold signal in the TPC. This allowed continuous {\it in situ} monitoring of the yield $Y_Z$ of these events. The average values for $Y_Z$ over the three data sets are given in 
Table~\ref{tbl:RunTable} and track well with the humidity sensor readings. 
To calibrate the necessary correction, 
special runs were conducted in which the hydrogen gas was doped 
with known amounts of nitrogen or water vapor.
The changes in the disappearance rate and $Y_Z$ 
were measured relative to the pure, undoped hydrogen data.
Scaling by the observed $Y_Z$ then determined the corrections for residual 
impurities: $(-7.80\pm1.87)$\,\is and $(-4.54\pm 0.93)$\,\is in R06 and
R07, respectively.

To obtain the final muon disappearance rate, the  muon stops
and electron tracks were first sorted into muon-electron
pairs. The decay time, $t\equiv t_{\textrm{eSC}} - t_{\mu\textrm{SC}}$, was histogrammed and fit with the function $N(t) = N_0 w \lambda_{\mu^{-}} e^{-\lambda^{}_{\mu^{-}} t}+B$ over the range 160~ns $< t <$ 19000~ns. The bin
width $w$ was fixed at 80~ns, while $N_0$, $B$ and
$\lambda_{\mu^{-}}$ were free parameters. To avoid analysis bias, the
exact clock frequency was hidden from the analyzers.
After it was revealed, we obtained 
\begin{align}
\label{eq:lambda_minus}
\lambda_{\mu^{-}} (\textrm{R}06 )  & = 455\,857.3 \pm 7.7_{\textrm{stat}} \pm 5.1_{\textrm{syst}}~\mathis , \\
\lambda_{\mu^{-}} (\textrm{R}07 )  & =       455\,853.1 \pm 8.3_{\textrm{stat}} \pm 3.9_{\textrm{syst}}~\mathis .
\end{align}
Because the fit $\chi^{2}/\textrm{DOF} = 1.2\pm0.1$ was slightly larger than 
expected for both R06 and R07, these values reflect inflated statistical uncertainties 
following the $S$-factor prescription~\cite{Beringer:1900zz}. The three-parameter 
fitting procedure was complemented by applying a full kinetics fit which 
included all atomic- and molecular-state effects as well as water and 
nitrogen impurities; the result was consistent within 0.2\,\isn.

In order to check the consistency of our result, we examined changes in
$\lambda_{\mu^{-}}$ with respect to variations in data selection. 
The fit start and stop times were varied over a
range of several microseconds and the parameters remained
stable. Only statistical variations were observed when the data were sorted
chronologically by run number. Since many of the subtle couplings between the muon and electron definitions are geometrical, the observed stability of the result with respect to azimuth was a critical cross-check. 

\begin{table}[ht]
  \centering
  \caption{Applied corrections and systematic errors. \label{tbl:ErrorTable}}
  \begin{tabular}{lcrr}
    \hline
    Effect & ~~ & \multicolumn{2}{c}{ Corrections and uncertainties [\isn]}\\  
           & &  R06      &  R07 \\
    \hline\hline
    $Z>1$ impurities               & & $-7.8 \pm 1.87$ & $-4.54 \pm 0.93$ \\
    $\mu$-$p$ scatter removal             & & $-12.4 \pm 3.22$ & $-7.2 \pm 1.25$ \\
    $\mu p$ diffusion                & & $-3.1 \pm 0.10$ & $-3.0 \pm 0.10$ \\
    $\mu d$ diffusion                & & $ ~ \pm 0.74$ & $  ~ \pm 0.12$ \\
    Fiducial volume cut             & & $  ~ \pm 3.00$ & $   ~ \pm 3.00$   \\
    Entrance counter ineff.     & & $  ~ \pm 0.50$ & $   ~ \pm 0.50$   \\
    Electron track def.             & & $  ~ \pm 1.80$ & $   ~ \pm 1.80$   \\
\hline 
   Total $\lambda_{\mu^{-}}$ corr.   & & $ -23.30 \pm 5.20 $ & $-14.74 \pm 3.88$ \\ 
\hline \hline
    $\mu p$ bound state: $\Delta\lambda_{\mu p}$       & & $-12.3 \pm 0.00$ & $-12.3 \pm 0.00$ \\
    $pp\mu$ states: $\Delta\Lambda_{pp\mu}$       & & $-17.73 \pm 1.87$ & $-17.72 \pm 1.87$ \\
    \hline
  \end{tabular}
\end{table}
Table \ref{tbl:ErrorTable} summarizes the aforementioned corrections to our
$\lambda_{\mu^-}$ result as well as the systematic uncertainties. Two 
additional corrections are required to correctly express $ \lambda_{\mu^{-}}$ as:
\begin{equation}
\label{eq:fullCorrection}
\lambda_{\mu^{-}} = \left( \lambda_{\mu^{+}} + \Delta\lambda_{\mu p} \right) + \Lambda_{S} + \Delta\Lambda_{pp\mu}.
\end{equation}
Here $\Delta \lambda_{\mu p}$ is a calculable $\mu p$ bound-state 
effect \cite{Uberall:1960, VonBaeyer:1979yg}, while $\Delta\Lambda_{pp\mu}$ 
accounts for the around 3\% of muons that capture from molecular states. The latter depends on $\lambda_{op}$ and $\lambda_{pp\mu}$ and is derived from fits to simulated data generated with the precise experimental conditions (gas density $\phi = 0.0115 \pm 0.0001$, 
background level, impurity concentrations, and fit ranges).
We used $\lambda_{op} = (6.6 \pm 3.4) \times
10^{4}$~\cite{Kammel:2010} and a newly determined value of  $\lambda_{pp\mu}=
(1.94 \pm 0.06)\times 10^6$\,\isn~\cite{knaack}, which was
measured by admixing $19.6\pm1.1$~ppm of
argon to the TPC's hydrogen. It agrees with the previous world
average~\cite{Kammel:2010} but is three times more precise.
Evaluating Eq.~\ref{eq:fullCorrection} and including the updated 
positive muon decay rate of $\LambdaMuplus \pm \LambdaMuplusErr$~\isn~\cite{Beringer:1900zz, Webber:2011}, we determine the singlet capture rates:
\begin{align}
\label{eq:LambaSMucap}
\Lambda_{S} (\textrm{R06} )  & = \LambdaSTen \pm \LambdaSStatTen_{\textrm{stat}} \pm \LambdaSSystTen_{\textrm{syst}}~\mathis , \\
\Lambda_{S} (\textrm{R07} )  & = \LambdaSEleven \pm \LambdaSStatEleven_{\textrm{stat}} \pm \LambdaSSystEleven_{\textrm{syst}}~\mathis . 
\end{align}
We also update slightly our previous publication~\cite{Andreev:2007}
using the latest values for $\lambda_{\mu^{+}}$, $\lambda_{op}$, and
$\lambda_{pp\mu}$, to obtain
$\Lambda_{S} (\textrm{R04}) = \LambdaSEight \pm \LambdaSStatEight_{\textrm{stat}} \pm \LambdaSSystEight_{\textrm{syst}}~\mathis .$
Accounting  for correlated systematics among these three data sets, we 
report a final, combined result
\begin{equation}
\label{eq:LambdaSTotal}
\Lambda_{S}^{\textrm{MuCap}} = \LambdaS \pm \LambdaSStat_{\textrm{stat}} \pm \LambdaSSyst_{\textrm{syst}}~\mathis .
\end{equation}
This new result is in excellent agreement with recent theory~\cite{Bernard:2000et, Ando:2000zw, Czarnecki:2007}. From the latest calculation~\cite{Czarnecki:2007},
we derive 
\begin{equation}
\begin{split}
\Lambda^{\textrm{Th}}_S&(g^{}_A, g^{}_P)= (712.7 \pm 3.0 \pm  \LambdaSThErrRad) \times \\
& \big[1 + 0.6265 (g^{}_A -g_A^{\textrm{PDG}}) \\ 
&  - 0.0108(g^{}_P - g_{P}^{\textrm{Th}} )\big]^{2} ~\mathis,
\end{split}
\end{equation}
where all form factors are evaluated at $q_0^2$. Equation (9) quantifies the dependence of the theoretical capture rate on
the choice of $g^{}_P$, relative to value $g^{\textrm{Th}}_P = 8.2$ used in Ref.~\cite{Czarnecki:2007},  and
on $g^{}_A$, relative to the latest $g_A^{\textrm{PDG}}(0)=1.2701\pm 0.0025$~\cite{Beringer:1900zz}. The two uncertainties in the equation
stem from limited knowledge of $g^{}_A$ and radiative corrections. 
Setting $\Lambda^{\textrm{Th}}_S(g_A^{\textrm{PDG}},
g_P^{\textrm{MuCap}})$ to $\Lambda_{S}^{\textrm{MuCap}} $ gives
\begin{equation}
\label{eq:gpmucap}
g_{P}^{\textrm{MuCap}} (q_0^2 = -0.88\,m_\mu^2)=8.06 \pm 0.48 \pm 0.28,
\end{equation}
where the two uncertainties arise from the error propagation of
$\Lambda_{S}^{\textrm{MuCap}}$ and
$\Lambda_{S}^{\textrm{Th}}$, respectively. 
If we would have updated $g^{}_A(0)$ to 1.275, as
advocated in~\cite{Blucher:2012} and supported by recent
measurements of the neutron $\beta$-decay asymmetry~\cite{Mund:2012fq, Mendenhall:2012tz}, the $g^{}_P$
extracted from MuCap would have increased to 8.34.

\begin{figure}
\begin{center}
  \includegraphics[angle=90, width=0.48\textwidth]{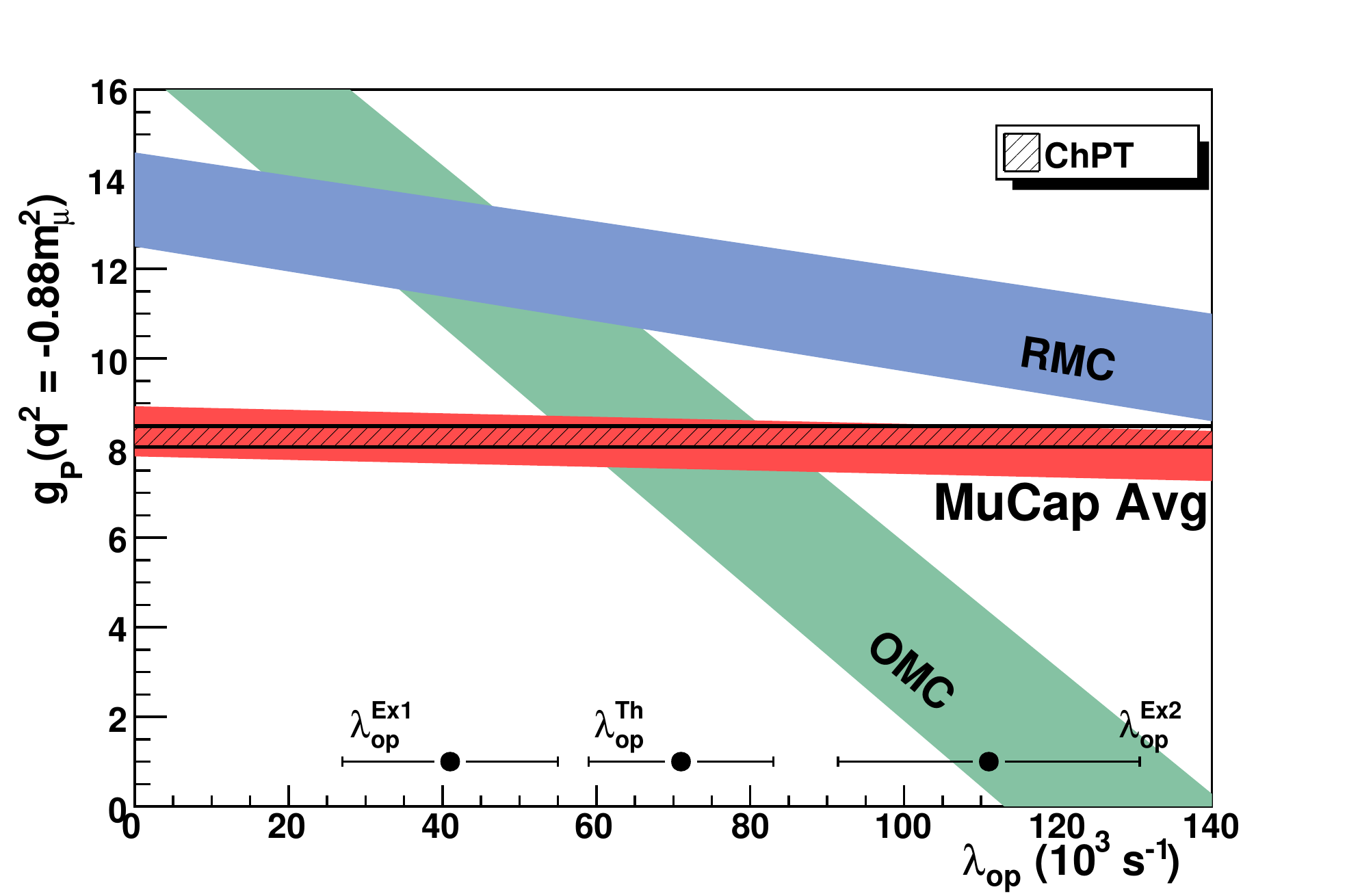}
  \caption{Extracted values for \gp\ as a function of the poorly known molecular
    transition rate $\lambda_{op}$ \cite{Bardin:1981cq,Clark:2005as,Bakalov:1980fm}.  In contrast to earlier experiments
    (OMC \cite{Bardin:1980mi}, RMC \cite{Wright:1998gi}), MuCap is rather insensitive to this parameter. }
  \label{lop.fig}
\end{center}
\end{figure}

Figure \ref{lop.fig} illustrates the excellent agreement with the theoretical prediction, Eq.~(\ref{equation:gp_theory}), and highlights MuCap's reduced sensitivity to the molecular parameter $\lambda_{op}$. This answers the long-standing challenge of an unambiguous
measurement of \gp, generated by the mutual inconsistency of earlier experiments (OMC, RMC) and their strong sensitivity to $\lambda_{op}$. Corroborating values for $g_P$ are obtained in recent analyses~\cite{PhysRevLett.108.052502, Gazit:2009ye} of an earlier 0.3\% measurement of muon capture on $^3$He~\cite{Ackerbauer:1997rs}, with uncertainties limited by theory. MuCap provides the most precise determination of \gp\ in the theoretically clean $\mu p$ atom and verifies a fundamental prediction of low-energy QCD.

We are grateful to the technical staff of the collaborating institutions, in particular of the host laboratory PSI. We thank M. Barnes, G. Wait, and A. Gafarov for the design and development of the kicker, the Demon collaboration for providing neutron detectors, the AMS team at the ETH Z\"urich for the deuterium measurements, and A. Adamczak, N. Bondar, D.B.~Chitwood, P.T. Debevec, T. Ferguson, J. Govaerts, S.~Kizilgul, M. Levchenko, and C.S.~{\"O}zben for their contributions.  
This work was supported in part by the U.S. NSF, the U.S. DOE and CRDF, PSI, the Russian Academy of Sciences and the Grants of the President of the Russian Federation. NCSA provided essential computing resources. 

\bibliography{MuCap_PRL}

\end{document}